\documentclass[prl,twocolumn,amsmath,amssymb,superscriptaddress,longbibliography]{revtex4-1}
\usepackage{graphicx,amsmath,relsize,epstopdf,upgreek,color,mathtools,bm,mathptmx}
\usepackage[colorlinks=true,linkcolor=blue,citecolor=blue,urlcolor =blue]{hyperref}

\newcommand{\sugg}[1]{\textcolor{black}{#1}}
\newcommand{\suggtwo}[1]{\textcolor{black}{#1}}

\newcommand{\rvec}[1]{\pmb{#1}}
\newcommand{\dyadic}[1]{\pmb{#1}}

\newcommand{\D}{\mathrm{d}}

\newcommand{\TP}[1]{{#1}^\mathrm{\,\textsc{t}}}

\newcommand{\ML}[1]{\widehat{#1}_\textsc{ml}}

\newcommand{\FML}{\dyadic{F}_\textsc{ml}}
\newcommand{\gML}{\bm{g}_\textsc{ml}}

\newcommand{\tr}[1]{\mathrm{tr}\!\left\{#1\right\}}
\newcommand{\Tr}[1]{\mathrm{Tr}\!\left\{#1\right\}}

\allowdisplaybreaks
\begin{document}

\title{Probing Bayesian credible regions intrinsically: a feasible error certification for physical systems}

\author{C.~Oh}
\affiliation{Center for Macroscopic Quantum Control, Department of Physics and Astronomy, Seoul National University, 08826 Seoul, South Korea}

\author{Y.~S.~Teo}
\email{ys\_teo@snu.ac.kr}
\affiliation{Center for Macroscopic Quantum Control, Department of Physics and Astronomy, Seoul National University, 08826 Seoul, South Korea}
\affiliation{Frontier Physics Research Division, Department of Physics and Astronomy, Seoul National University, 08826 Seoul, South Korea}

\author{H.~Jeong}
\affiliation{Center for Macroscopic Quantum Control, Department of Physics and Astronomy, Seoul National University, 08826 Seoul, South Korea}

      \begin{abstract}
        Standard computation of size and credibility of a Bayesian credible region for certifying any point estimator of an unknown parameter (such as a quantum state, channel, phase, \emph{etc.}) requires selecting points that are in the region from a finite parameter-space sample, which is infeasible for a large dataset or dimension as the region would then be extremely small. We solve this problem by introducing the in-region sampling theory to compute both region qualities just by sampling appropriate functions over the region itself using any Monte~Carlo sampling method. We take in-region sampling to the next level by \sugg{understanding the credible-region capacity (an alternative description for the region content to size) as} the average $l_p$-norm distance $(p>0)$ between a random region point and the estimator, and present analytical formulas for $p=2$ to estimate \sugg{both the capacity} and credibility for any dimension and sufficiently large dataset \sugg{without Monte~Carlo sampling, thereby providing a quick alternative to Bayesian certification.} All results are discussed in the context of quantum-state tomography.
      \end{abstract}

\maketitle

\emph{Introduction.---}Parameter reconstruction from datasets is a preliminary task in the study of natural sciences. In quantum theory, proper reconstruction of quantum states~\cite{Smithey:1993uq,Chuang:2000fk,Rehacek:2007ml,Teo:2011me,Zhu:2014aa}, quantum channels~\cite{OBrien2004aa,Teo2011aa,Fiurasek2015aa,Varga2018aa}, interferometric phases~\cite{Caves:1981aa,Dorner:2009aa}, \emph{etc.}, is the root to successful executions of all quantum-information protocols~\cite{Ladd:2010aa,D-Dobrzanski:2016aa,Campbell:2017aa,Ladd:2010aa,Lekitsch:2017aa}. A parameter estimator must be accompanied by an appropriate error certification to ascertain its reliability for future physical predictions. Bootstrapping or resampling~\cite{Efron:1993bs,Davison:1997ri}, which generates mock data from collected ones to obtain ``error-bars'', can result in highly overoptimistic ``error-bar'' lengths~\cite{Suess:2017np} that do not accurately characterize the estimator. From the principles of hypothesis testing, one can instead construct Bayesian \emph{credible regions}~\cite{Shang:2013cc,Li:2016da} based on the collected data. These credible regions are distinct from the frequentists' confidence regions~\cite{Christandl:2012qs,Blume-Kohout:2012eb,Faist:2016eb}, which are constructed from the complete (often assumed) distribution of estimators that includes all unobserved ones in the experiment.

A credible region $\mathcal{R}$, which is a Bayesian error region constructed from experimentally observed data $\mathbb{D}$, requires the specification of its \emph{size} and \emph{credibility}, which is the probability that the true parameter is inside $\mathcal{R}$. It is well-known from \cite{Shang:2013cc} that the latter is readily derived so long as the functional behavior of the former with the shape of $\mathcal{R}$ is known. \suggtwo{As the size of $\mathcal{R}$ is defined as the volume fraction of the full parameter space $\mathcal{R}_0$, its computation conventionally requires one to first obtain a large sample of points in $\mathcal{R}_0$, and later discard (usually very many) points that are outside $\mathcal{R}$.} Acquiring a sufficiently large sample of $\mathcal{R}_0$ for a subsequently accurate \sugg{sample filtering} is doable with a number of Monte~Carlo (MC) methods~\cite{Shang:2015mc,Seah:2015mc}, most notably the Hamiltonian Markov-chain MC, \emph{provided that} $\mathcal{R}$ is not small. In practice, however, when data sample-size $N$ becomes even moderately large, the region $\mathcal{R}$ (of size $\sim N^{-d/2}$~\cite{Teo:2018aa} for a $d$-dimensional parameter) is too tiny for any \sugg{MC-filtering} sampling to be practically feasible. In \cite{Teo:2018aa,Oh:2018aa}, closed-form approximations are given to estimate both region qualities for large $N$ without \sugg{MC-filtering}, with the premise that the volume of $\mathcal{R}_0$ is known.

In this Letter, we develop an \emph{in-region sampling} theory to compute the size and credibility with neither \sugg{MC-filtering} from nor any geometrical knowledge about $\mathcal{R}_0$ (such as its volume). We first prove the central lemma which states that both region qualities are computable from the average of log-likelihood over $\mathcal{R}$. We next discuss the hit-and-run MC algorithm~\cite{Belisle:1993aa,Smith:1996aa,Lovasz:2006aa,Kiatsupaibul:2011aa} \sugg{as one of the many numerical tools} to perform direct region-average computation. As a strategic bonus, we make use of the region-average \suggtwo{concept} in in-region sampling to define the region capacity of $\mathcal{R}$ induced by an $l_p$-norm ($p>0$) between two points in $\mathcal{R}$. This would allow us to derive fully operational asymptotic approximation formulas for $p=2$ (squared-error metric) to carry out rapid error certifications without numerical computations. All results are demonstrated and verified for multi-qubit tomography.

\begin{figure}[t]
	\center
	\includegraphics[width=0.8\columnwidth]{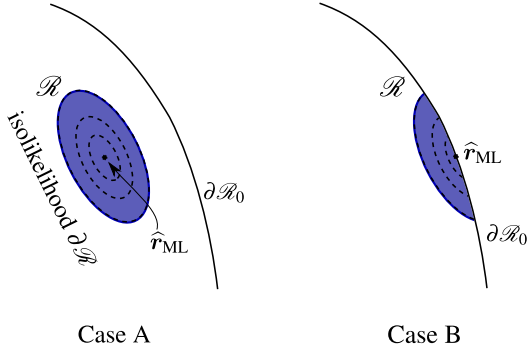}
	\caption{\label{fig:regions}(Color Online) Since any relevant $\lambda$ value that gives a reasonably large credibility $C_\lambda<1$ typically yields a small CR $\mathcal{R}_\lambda$, there exist only two general cases. Case~A refers to the situation where $\ML{\rvec{r}}$ is an interior point of $\mathcal{R}_0$, and Case~B refers to that where $\ML{\rvec{r}}$ is on the boundary $\partial\mathcal{R}\cap\partial\mathcal{R}_0$. It is easy to determine which is the case for $\ML{\rvec{r}}$. In quantum-state tomography, for instance, this would correspond to checking if the state estimator is rank-deficient.}
\end{figure}

{\it \sugg{Error-region} size and credibility.---}For a given \emph{informationally complete} (IC) dataset $\mathbb{D}$, we would like to reconstruct the unknown $d$-dimensional parameter $\rvec{r}$ (\sugg{vectorial in general}) that fully characterizes some physical system. We shall assume that the parameter space $\mathcal{R}_0$ (of quantum states, channels, Cartesian-product of independent quantities, \emph{etc.}) for the physical system of interest is convex, and take the unique \emph{estimator} $\widehat{\rvec{r}}=\ML{\rvec{r}}$ to be the \emph{maximum-likelihood} (ML) estimator~\cite{Aldrich:1997ml,Rehacek:2007ml,Teo:2015qs}, that is the estimator that maximizes the \emph{likelihood} $L=L(\mathbb{D}|\rvec{r}')$. It was formally shown in \cite{Shang:2013cc} that the optimal Bayesian credible region (CR) $\mathcal{R}$ for $\ML{\rvec{r}}$ has an isolikelihood boundary $\partial\mathcal{R}$---a boundary of constant likelihood---and every interior point possessing a likelihood $L\geq\lambda\,L_\text{max}$ (see Fig.~\ref{fig:regions}). Its size and credibility are
\begin{align}
S_\lambda\equiv&\,\int_{\mathcal{R}_\lambda}(\D\,\rvec{r}')=\int_{\mathcal{R}_0}(\D\,\rvec{r}')\,\eta(L-\lambda L_\text{max})\,,\nonumber\\
C_\lambda\equiv&\,\int_{\mathcal{R}_\lambda}(\D\,\rvec{r}')\,L/L(\mathbb{D})=\int_{\mathcal{R}_0}(\D\,\rvec{r}')\,\eta(L-\lambda L_\text{max})\,L/L(\mathbb{D})\,,
\end{align}
where the volume measure $(\D\,{\rvec{r}})$ incorporates some prescribed prior distribution \sugg{$p(\rvec{r})$}, $\eta$ is the Heaviside function, $L(\mathbb{D})=\int_{\mathcal{R}_0}(\D\,\rvec{r}')\,L(\mathbb{D}|\rvec{r}')$, and $0\leq\lambda\leq1$ characterizes the shape and size of $\mathcal{R}_\lambda$, so that $\mathcal{R}_{\lambda=0}=\mathcal{R}_0$ and $\mathcal{R}_{\lambda=1}=\{\ML{\rvec{r}}\}$. Hence, $S_\lambda$ measures the total prior content of $\mathcal{R}_\lambda$ that monotonically decreases with increasing $\lambda$, and $C_\lambda$ its posterior content that expresses the probability that $\rvec{r}\in\mathcal{R}_\lambda$. Both $C_{\lambda=\lambda_0}$ (pre-chosen to be 0.95 say) and the corresponding $S_{\lambda=\lambda_0}$ are reported together with $\ML{\rvec{r}}$. The relation
\begin{equation}
C_\lambda=\left(\lambda S_\lambda+\int^1_\lambda\D\,\lambda' S_{\lambda'}\right)\bigg/\int^1_0\D\,\lambda' S_{\lambda'}
\label{eq:Cpr}
\end{equation}
means that a single $\rvec{r}'$-integration for $S_\lambda$ is sufficient to acquire $C_\lambda$~\cite{Shang:2013cc}. In realistic experiments, where the desired number of data copies $N<\infty$ is usually large (which we assume unless otherwise stated), the likelihood $L$ becomes a Gaussian function owing to \suggtwo{the} central limit theorem and peaks strongly around $\ML{\rvec{r}}$. In this case, $S_\lambda$ becomes very small even for small $\lambda$ or large $C_\lambda$ (the desired situation). Therefore, MC-filtering produces almost no yield as such a finite sample would surely miss $\mathcal{R}_\lambda$ for a reasonably high $C_\lambda$.

\suggtwo{We inform that one systematic guide to report error regions is to invoke the elegant notion of evidence, which leads to the so-called plausible region~\cite{Li:2016da,Evans:2016aa,Al-Labadi:2018aa,Teo:2018aa,Oh:2018aa} for $\mathbb{D}$, in which all points have posterior probabilities larger than or equal to their prior probabilities---a physical measure of statistical significance. Then $C_\lambda$ should \emph{not} exceed the credibility of this plausible region in order for the CR to contain only plausible points (refer to our companion article~\cite{PRA} for details).}

{\it In-region sampling theory.---}We shall now propose a way to compute \emph{both} $S_\lambda$ and $C_\lambda$ without MC-filtering. The physical intuition behind our theory is to realize that if one inspects the average of some quantity $q_\lambda$ over the region $\mathcal{R}_\lambda$ [formally denoted by $\overline{q_\lambda}^{\mathcal{R}_\lambda}=\int_{\mathcal{R}_\lambda}(\D\rvec{r}')q_\lambda(\rvec{r}')/\int_{\mathcal{R}_\lambda}(\D\rvec{r}')$], then its rate of change with $\lambda$ actually encodes information about the behavior of $S_\lambda$ with $\lambda$. A shrinkage of $\mathcal{R}_\lambda$, for example, translates to an exclusion of some $q_\lambda$ values from the region-average. More precisely, this leads to the\\

\noindent
{\bf Region-average computation (RAC) lemma}: \emph{For any prior $(\D\,\rvec{r}')$ and $N$, the prior content $S_\lambda$ (up to a multiplicative factor), and hence the credibility $C_\lambda$, are all inferable from the $\mathcal{R}$-average quantity $u_\lambda=\overline{\log L(\mathbb{D}|\rvec{r}')-\log(\lambda\,L_\text{max})}^{\mathcal{R}_\lambda}$.}\\[1ex]

We prove this lemma by taking the first-order derivative of $u_\lambda S_\lambda$ in $\lambda$. Upon noting that $\partial S_\lambda/\partial\lambda=\int_{\mathcal{R}_0}(\D\,\rvec{r}')\,\delta(L-\lambda\,L_\text{max})$, we end up with the following first-order differential equation
\begin{equation}
\dfrac{\partial y_\lambda}{\partial\lambda}=-\dfrac{y_\lambda}{\lambda\,u_\lambda}
\label{eq:ode}
\end{equation}
that characterizes the full evolution of $y_\lambda=u_\lambda\,S_\lambda$ given the boundary value $S_{\lambda=0}=1$. Equation~\eqref{eq:ode} can be solved easily by iterating $y_{\lambda_{j+1}}=y_{\lambda_j}-y_{\lambda_j}/(\lambda_j u_{\lambda_j})$ following Euler's method~\cite{Butcher:2003aa}, so that $C_\lambda$ can thereafter be computed using Eq.~\eqref{eq:Cpr}. This closes our constructive proof of the RAC lemma. 

For any prior distribution $p(\rvec{r})$, there exist many MC~\cite{Shang:2015mc,DelMoral:2006aa} schemes to compute $u_\lambda$, many of which use Markov-chain algorithms. \emph{Hit-and-run} sampling~\cite{Belisle:1993aa,Smith:1996aa,Lovasz:2006aa,Kiatsupaibul:2011aa} is one such extensively-studied scheme. The mechanism behind hit-and-run starts with the construction of a simple finite convex set $\mathcal{B}\supseteq\mathcal{R}_\lambda$. For $N\gg1$ and some $\lambda$, two general cases exist as shown in Fig.~\ref{fig:regions}. In Case~A, we define $\mathcal{B}$ as the hyperellipsoid $\mathcal{E}_\lambda$ centered at $\rvec{r}_\text{c}=\ML{\rvec{r}}$ that profiles the Gaussian $L$ whenever $\ML{\rvec{r}}$ is an interior point. In Case~B, where $\ML{\rvec{r}}$ is a boundary point on $\partial\mathcal{R}_0$, we set $\mathcal{B}$ as the (truncated) hyperellipsoid $\mathcal{E}'_\lambda$ centered at $\rvec{r}_\text{c}=\ML{\rvec{r}}+\FML^{-1}\,\gML$, where $\FML$ is the Fisher information evaluated at $\ML{\rvec{r}}$ and $\gML=\partial\log L/\partial\rvec{r}'|_{\rvec{r}'=\ML{\rvec{r}}}$. Next, starting from a reference point in $\mathcal{R}_\lambda$, say the ML estimator $\ML{\rvec{r}}$, a finite line segment, with endpoints on $\partial\mathcal{B}$, passing through this point is generated and a random point is picked repeatedly along this line until it lies in $\mathcal{R}_\lambda$, thereafter becoming the next reference through which a new finite line segment is generated to find the next point in $\mathcal{R}_\lambda$. The final $\mathcal{R}_\lambda$ sample is then used to compute \emph{any} $\mathcal{R}_\lambda$-average quantity. \suggtwo{The key point is that a hyperellipsoidal $\mathcal{B}$ for hit-and-run is constructed based on the central limit theorem, where the $N\gg1$ condition guarantees that the physical region is asymptotically contained in $\mathcal{B}$. To play it safe, a good idea would be to choose a hyperellipsoid that is, say, twice the size of the supposed one given by the theorem.}

Beginning with $k=1$ and $\rvec{r}_\text{ref}=\ML{\rvec{r}}$ of $N\gg1$, the accelerated version of hit-and-run~\cite{Belisle:1993aa,Smith:1996aa,Kiatsupaibul:2011aa} for any given prior distribution $p(\rvec{r})$ runs as follows: {\bf1.}~Generate a random line segment characterized by $\rvec{y}=\rvec{r}_\text{ref}+\mu\,\rvec{\mathrm{e}_v}$, where $\rvec{\mathrm{e}_v}=\rvec{v}/|\rvec{v}|$ and $\rvec{v}$ follows the standard Gaussian distribution (mean 0 and variance 1 for each column entry). Its endpoints are parametrized by $\mu_\pm=[-b\pm\sqrt{b^2-a(c-1)}]/a$, where $\rvec{\Delta}=\rvec{r}_\text{ref}-\rvec{r}_\text{c}$, $a=\TP{\rvec{\mathrm{e}_v}}\dyadic{A}\,\rvec{\mathrm{e}_v}$, $b=\TP{\rvec{\Delta}}\dyadic{A}\,\rvec{\mathrm{e}_v}$, $c=\TP{\rvec{\Delta}}\dyadic{A}\,\rvec{\Delta}$, $\dyadic{A}=\FML/(-2\log\lambda')$, $2\log(\lambda/\lambda')=\gML^\textsc{t}\FML^{-1}\gML$ [$\gML=\rvec{0}$ and $\lambda'=\lambda$ for Case~A].	{\bf2.}~Define $\beta_1\equiv\mu_\text{min}=\min\{\mu_+,\mu_-\}$ and $\beta_2\equiv\mu_\text{max}=\max\{\mu_+,\mu_-\}$. {\bf3.}~Pick a random number $\beta_1\leq\beta\leq\beta_2$ according to the marginal probability distribution $p(\rvec{r}_\text{ref}+\beta\,\rvec{\mathrm{e}_v})/\int\D\beta'p(\rvec{r}_\text{ref}+\beta'\,\rvec{\mathrm{e}_v})$ truncated in the interval [$\beta_1,\beta_2$] and obtain $\rvec{r}_\text{test}=\rvec{r}_\text{ref}+\beta\,\rvec{\mathrm{e}_v}$. {\bf4.}~Determine whether $\rvec{r}_\text{test}\in\mathcal{R}_\lambda$. If so, define $\rvec{r}_\text{ref}=\rvec{r}_\text{test}$, raise $k$ by 1, and go to {\bf1}. If not, set $\beta_1=\beta$ if $\beta<0$ or $\beta_2=\beta$ if $\beta>0$, and repeat {\bf3} and {\bf4}. Sampling terminates when $k>K_\text{smp}$ for a prechosen $K_\text{smp}$.

We emphasize that the Gaussian approximation serves only as an efficient guide to contain the sampling space. An additional criterion that $\log L>\log(\lambda L_\text{max})$ may be used to further ensure that all sampled points truly lie in $\mathcal{R}_\lambda$, although this is almost always the case for $N\gg1$. One main technical issue for Markov-chain schemes is that the convergence rate is strongly dependent on the starting point (finite sample-point correlation). It is well-known, however, that hit-and-run converges fast to $p(\rvec{r})$ (with essentially polynomial complexity) so long as it starts from \emph{any} interior point. As an example in 4-qubit tomography, such an interior point can be generated in about 10 seconds per $\lambda$ with $N>2\times10^6$ and 4096 measurement outcomes using accelerated projected gradient method~\cite{Shang:2017sf} to minimize the function $[1-(\rvec{x}-\rvec{r}_\text{c})\bm{\cdot}\dyadic{A}\bm{\cdot}(\rvec{x}-\rvec{r}_\text{c})]^2$ 32 times~(see for instance \cite{Lovasz:2006aa,Lovasz:1999aa} and our companion article~\cite{PRA} for more technical discussions).

{\it \sugg{Region capacity}.---}The region-average methodology used to feasibly compute $S_\lambda$ (and $C_\lambda$) invites more options to gauge the capacity of $\mathcal{R}$. Instead of measuring prior contents, we may check how close is a randomly-chosen point in $\mathcal{R}$ from $\ML{\rvec{r}}$ on average. Formally, the $\mathcal{R}$-average
\begin{equation}
S_{\mathcal{D},\lambda}\equiv\overline{\mathcal{D}(\rvec{r}',\ML{\rvec{r}})}^{\mathcal{R}_\lambda}=\int_{\mathcal{R}_\lambda}(\D\,\rvec{r}')\,\mathcal{D}(\rvec{r}',\ML{\rvec{r}})\bigg/\int_{\mathcal{R}_\lambda}(\D\,\rvec{r}')
\label{eq:Slbd}
\end{equation}
for the \sugg{capacity} of $\mathcal{R}_\lambda$ now depends additionally on the metric $\mathcal{D}(\rvec{r}',\ML{\rvec{r}})$ one chooses to measure this average distance.

\begin{figure}[t]
	\center
	\includegraphics[width=0.9\columnwidth]{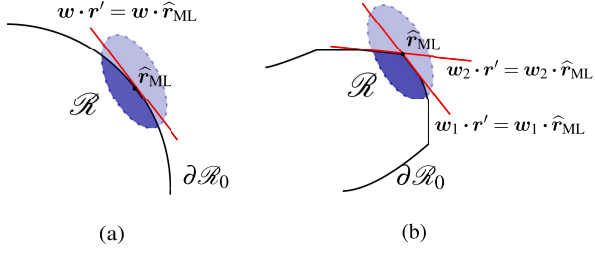}
	\caption{\label{fig:bd}(Color Online) Barring unforeseen pathological examples, we shall assume that the $\mathcal{R}_0$ for any physical system possesses a boundary $\partial\mathcal{R}_0$ that is either (a) a smooth surface, or (b) has corners and edges. For the latter, a corner at which an ML estimator might reside can be well approximated by multiple hyperplanes if $N\gg1$.}
\end{figure}

One can argue that if the metric is an $l_p$-norm of $p>0$, $S_{\mathcal{D},\lambda}$ monotonically decreases with $\lambda$ when $N\gg1$ for an appropriate $(\D\,\rvec{r}')$. To see this we begin with  $\mathcal{D}\equiv\mathcal{D}_p(\rvec{r}',\ML{\rvec{r}})=\left(\sum_j|r'_j-\widehat{r}_{\textsc{ml},j}|^p\right)^{1/p}$. According to Fig.~\ref{fig:bd}, after the substitution $\rvec{r}''=\rvec{r}'-\ML{\rvec{r}}$, we have for the more complicated Case~B,
\begin{align}
S_{\mathcal{D}_p,\lambda}\rightarrow&\,\dfrac{\displaystyle\int(\D\,\rvec{r}'')\,\mathcal{D}_p\,\eta(1-\TP{\rvec{r}''}\FML\rvec{r}''/(-2\log\lambda))\prod_j\eta(\rvec{w}^\textsc{t}_j\rvec{r}'')}{\displaystyle\int(\D\,\rvec{r}'')\eta(1-\TP{\rvec{r}''}\FML\rvec{r}''/(-2\log\lambda))\prod_j\eta(\rvec{w}^\textsc{t}_j\rvec{r}'')}\nonumber\\
\sim&\,\sqrt{-2\log\lambda}\quad\left[\text{if $(\D\,\alpha\,\rvec{r}'')=g(\alpha)\,(\D\,\rvec{r}'')$}\right]\,.
\end{align}
The same conclusion for Case~A follows by definition, and remains unchanged also for $\mathcal{D}_p(\rvec{r}',\ML{\rvec{r}})=\sum_j|r'_j-\widehat{r}_{\textsc{ml},j}|^p$ since $S_{\mathcal{D}_p,\lambda}\sim(-\log\lambda)^{p/2}$ is also monotonic in $\lambda$. These imply that $S_{\mathcal{D}_p,\lambda}$ induced by any $l_p$-norm behaves as a proper capacity measure in the limit $N\gg1$ under a sufficient class of priors that includes the uniform primitive prior. The new practice for Bayesian CR certification is then to report the three-tuple $\left(\ML{\rvec{r}},C_{\lambda_0}(=0.95 \text{ say}),S_{\mathcal{D}_p,\lambda_0}\right)$ for some $p>0$.

\begin{figure}[t]
	\center
	\includegraphics[width=0.95\columnwidth]{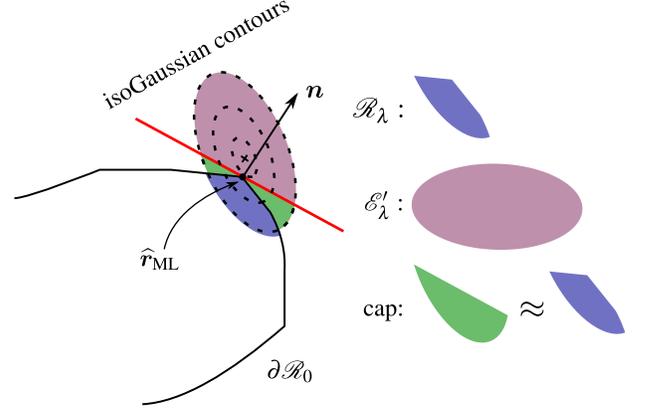}
	\caption{\label{fig:cap}(Color Online) After expanding the likelihood $L$ about $\ML{\rvec{r}}$ to a Gaussian function centered at $\rvec{r}_\text{c}$ (cross) with its own isoGaussian contours, a hyperplane (red solid line) is introduced in a manner that its normal $\rvec{n}$ is orthogonal to the isoGaussian curve at $\ML{\rvec{r}}$ to form a cap.}
\end{figure}

{\it \sugg{Analytical error certification with region capacity.}---}It turns out that the approximated extensions of all $\int_{\mathcal{R}_0}$ integrals to the whole $\rvec{r}'$ space free \emph{all} $\mathcal{R}$-average quantities from any geometrical dependence on $\mathcal{R}_0$, unlike $S_\lambda$ that asymptotically depends on $\mathcal{R}_0$'s volume~\cite{Teo:2018aa}. We may then use this observation to acquire asymptotic formulas for $S_{\mathcal{D}_p,\lambda}$ and $u_\lambda$ to perform approximate analytical error certifications. To this end, we regard $S_2\equiv S_{\mathcal{D}_2}$ induced by the squared $l_2$-norm ($p=2$), $\mathcal{D}\equiv|\rvec{r}'-\ML{\rvec{r}}|^2$, as the prototypical metric-induced capacity measure for $\mathcal{R}_\lambda$. Let us first discuss the case in which $\ML{\rvec{r}}$ is an interior point of $\mathcal{R}_\lambda$ (Case~A). Since $\mathcal{R}_\lambda=\mathcal{E}_\lambda$, finding $S_2$ becomes the business of doing a hyperellipsoidal average of $\mathcal{D}$. This gets us to
\sugg{\begin{equation}
S_{2\text{A},\lambda}=\Tr{\FML^{-1}}\dfrac{(-\log\lambda)}{d/2+1}\,,\quad
u_{\text{A},\lambda}=-\dfrac{2}{d+2}\log\lambda\,.
\label{eq:thA}
\end{equation}}
The logarithmic divergences in $\lambda$, a derivation byproduct from Gaussian approximation of $L$ and relaxation of $\partial\mathcal{R}_0$, pose no ill consequence so long as $N$ is sufficiently large such that $\mathcal{R}_\lambda\subset\mathcal{R}_0$ for all $\lambda$ values that give desirably large $C_\lambda<1$.

The situation becomes more complicated for Case~B, which demands geometrical knowledge about $\partial\mathcal{R}_0$ for an exact calculation of $S_2$~(see Fig.~\ref{fig:cap}). This tempts us to use a first-order approximation by expanding the likelihood $L$ about $\ML{\rvec{r}}$ to a Gaussian function of hyperellipsoidal-$\mathcal{E}'_\lambda$ profile centered at $\rvec{r}_\text{c}$, and next introducing a hyperplane containing $\ML{\rvec{r}}$ that is tangent to its isoGaussian (constant-Gaussian-value) contour. $S_2$ is then a hyperellipsoidal-cap (formed by the hyperplane and the hyperellipsoid from the Gaussian expansion of $L$) average. We refer the Reader to Sec.~VII of our companion article for all related technical calculations, and simply state the final formulas:
\begin{align}
S_{2\text{B},\lambda}=&\,2\,\Tr{\dyadic{M}}/\mathcal{N}_{d,l,1}\,,\nonumber\\
u_{\text{B},\lambda}=&\,[-\log \lambda'+\Tr{\gML\rvec{m}^\textsc{t}-\FML\,\dyadic{M}}/\mathcal{N}_{d,l,1}]\nonumber\\
&\,\times\log(\lambda\,L_\text{max})/\log(\lambda'\,L_\text{max})\,,
\label{eq:thB}
\end{align}
involving $V_d=\pi^{d/2}/(d/2)!$, $l=\sqrt{\log(\lambda/\lambda')/(-\log\lambda')}$, $\mathcal{N}_{d,l,x}=V_d\,\mathrm{I}_{(1-l)/2}((d+x)/2,(d+x)/2)$ depending on the incomplete beta function $\mathrm{I}_\cdot(\cdot,\cdot)$, and 
\begin{align}
\rvec{m}=&\,\left[-\dfrac{V_{d-1}}{l(d+1)}\left(1-l^2\right)^{(d+1)/2}+\mathcal{N}_{d,l,1}\right]\FML^{-1}\gML\,,\nonumber\\
\dyadic{M}=&\,\dfrac{-\log\lambda'}{d+2}\,\mathcal{N}_{d,l,3}\,\FML^{-1}+\dfrac{1}{2}\,\rvec{m}\,\gML^\textsc{t}\FML^{-1}\,.
\label{eq:auxop}
\end{align}
It is easy to see that Eqs.~\eqref{eq:thB} and \eqref{eq:auxop} include Case~A by recognizing that the ``effective $\lambda$'' ($\lambda'$) approaches $\lambda$ ($\gML=\rvec{0}$), so that $l\rightarrow0$ gives $\mathcal{N}_{d,0,x}=V_d$ and $\dyadic{M}=(-\log\lambda)\,\FML^{-1}/(d+2)$.

\begin{figure}[t]
	\center
	\includegraphics[width=1\columnwidth]{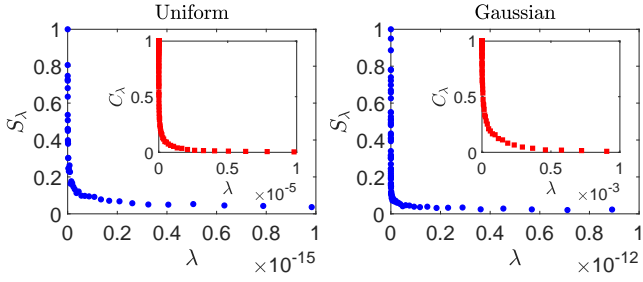}
	\caption{\label{fig:uC}(Color Online) Plots for $S_\lambda$ and $C_\lambda$ generated from the in-region sampling technique on three-qubit systems $(D=8)$, with a rank-1 $\ML{\rvec{r}}$, $M=512$ square-root measurement outcomes and $N/M=5000$. The rapidly decreasing $S_\lambda$ is a signature of typically small regions of such datasets, which cannot be handled with MC-filtering. The results for $C_\lambda$ obtained from the sampled $u_\lambda$ generated with 200 recursive steps of Euler's method to solve Eq.~\eqref{eq:ode} for $S_\lambda$. The flexibility of in-region sampling is demonstrated by presenting graphs sampled according to both the uniform and Gaussian distributions.}
\end{figure}

{\it Discussions for quantum-state tomography.---}All results presented thus far apply to arbitrary physical systems. Here, we specifically investigate quantum-state tomography, thereby endowing explicit forms to all important quantities that are pertinent to Bayesian CR error certification. 

For an unknown quantum state $\rho$ of Hilbert-space dimension $D$, every data-copy measurement in a tomography experiment is usually mutually independent, so that the log-likelihood $\log L=\sum^M_{j=1}n_j\log p_j$ catalogs the relative frequency data $\sum^M_{j=1}n_j=N$ of all $M$ measurement outcomes $\Pi_j\geq0$ $(\sum_j\Pi_j=1)$, each with the Born probability $p_j=\tr{\rho\Pi_j}$. We can express $\rho$ and $\Pi_j$ in terms of the Hermitian basis $\{1/\sqrt{D},\Omega_j\}^{D^2-1}_{j=1}$ such that $\tr{\Omega_j}=0$ and $\tr{\Omega_j\Omega_k}=\delta_{j,k}$, so that we may denote the ($d=D^2-1$)-dimensional $\rvec{r}=\tr{\rho\,\rvec{\Omega}}$ and $\rvec{q}_j=\tr{\Pi_j\,\rvec{\Omega}}$. This leads to $\FML=N\sum^M_{j=1}\rvec{q}_j\rvec{q}^\textsc{t}_j/p_{\textsc{ml},j}$ ($N\gg1$) and $\gML=\sum^M_{j=1}n_j\,\rvec{q}_j/p_{\textsc{ml},j}$ for the ML state estimator $\ML{\rho}$ of ML probabilities $p_{\textsc{ml},j}=\tr{\ML{\rho}\Pi_j}$. In concrete terms, for Case~A, $\ML{\rho}$ is full rank, such that the CR $\mathcal{R}_\lambda\approx\mathcal{E}_\lambda$; whereas for Case~B, $\ML{\rho}$ is rank-deficient and $\mathcal{R}_\lambda\approx\mathcal{R}_0\cap\mathcal{E}'_\lambda$ is therefore approximately a truncated $\mathcal{E}'_\lambda$ (covariance profile of the Gaussian expansion of $L$ about $\ML{\rvec{r}}$) by the quantum-state space $\mathcal{R}_0$---the convex set of unit-trace positive operators. The uniform $(\D\,\rvec{r}')$ is assumed.

\begin{figure}[t]
	\center
	\includegraphics[width=1\columnwidth]{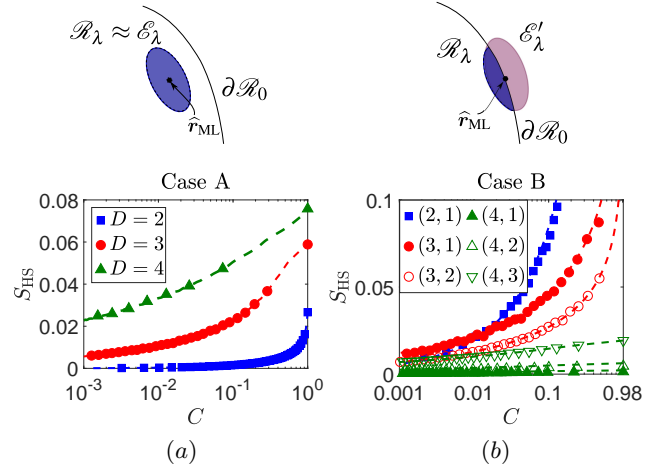}
	\caption{\label{fig:caseAB}(Color Online) The (magnified) per-$\mathbb{D}$ graphs of $S_\textsc{hs}$ versus $C$ for (a) Case~A and (b)~Case~B for various $D$, with $M=D^3$ random outcomes and $N/M=500$. The two-tuples in the legend of (b) represent $(D,\mathrm{rank}\{\ML{\rvec{r}}\})$. The respective dashed curves passing through the markers are calculated using Eqs.~\eqref{eq:thA} and \eqref{eq:thB}. The magnification factors (top to bottom, left to right in legend) for Case~A are 10, 50 and 150, and those for Case~B are 100, 200, 150, 10, 20 and 50.}
\end{figure}

To compare with the closed-form approximations in Eqs.~\eqref{eq:thA} and \eqref{eq:thB}, we pick the $l_2$-norm to measure the region capacity $S_\textsc{hs}\equiv S_2$ of $\mathcal{R}$, which is equivalent to the Hilbert-Schmidt (HS) distance for quantum states. We emphasize that for sufficiently large $N$, all arguments leading to the monotonicity of $S_{\mathcal{D},\lambda}$ still applies for Case~B as $\gML\rightarrow\rvec{0}$. Figures~\ref{fig:uC} and \ref{fig:caseAB} showcase our in-region sampling theory. The matches in both Case~A and B between theory and hit-and-run sampling are very good for moderate $D$, but are expected to have some discrepancies for more complex systems due to the more pronounced corners in $\partial\mathcal{R}_0$~\cite{Bengtsson:2013gm}. Instead, accelerated hit-and-run can be used, the complexity of which are analyzed in our companion article. 

{\it Conclusions.---}{In realistic multi-dimensional parameter estimation problems, sufficiently large dataset almost exclusively results in extremely small Bayesian credible regions relative to the entire parameter space. The conventional practice of first doing Monte~Carlo to sample the parameter space followed by sample filtering almost always fails to accurately construct such small error regions. Our technique of in-region sampling developed in this Letter is capable of constructing any such small regions efficiently with perfect yield. In-region sampling is equivalent to computing region-averages that is efficient with a wide range of numerical methods. The region-average perspective of in-region sampling allows us to operationally formulate an alternative concept of region capacity through averaging any $l_p$ distance norm between two credible-region points, for which, in the special case $p=2$, closed-form approximation formulas to facilitate ultrafast analytical Bayesian error estimations with sufficiently large datasets are readily available. Either way, efficient Bayesian error certifications can now be carried out on physical systems of varying complexity. For exceedingly large quantum systems where Monte~Carlo computations start to become visibly taxing, these asymptotic formulas can serve as large-scale approximate certifiers at least for high credibility values.

\begin{acknowledgments}
The authors thank J.~Shang for fruitful discussions, and acknowledge financial support from the BK21 Plus Program (21A20131111123) funded by the Ministry of Education (MOE, Korea) and National Research Foundation of Korea (NRF), the framework of international cooperation program managed by the NRF (NRF-2018K2A9A1A06069933), and the Basic Science Research Program through the NRF funded by the Ministry of Education (No. 2018R1D1A1B07048633).
\end{acknowledgments}

%



\end{document}